\documentclass[prb,superscriptaddress,twocolumn,showpacs]{revtex4-1}

\usepackage{graphicx}
\usepackage{amsmath,amssymb,bm}

\begin{document}

\renewcommand{\ss}[1]{_{\hbox{\tiny #1}}}

\title{Re-entrant spin-flop transition in nanomagnets}

\author{Mattia Crescioli}
\affiliation{Dipartimento di Fisica e Astronomia, Universit\`a di Firenze,
             via G.~Sansone 1, I-50019 Sesto Fiorentino, Italy}
\affiliation{Istituto dei Sistemi Complessi,
             Consiglio Nazionale delle Ricerche,
             via Madonna del Piano 10,
             I-50019 Sesto Fiorentino, Italy}
\author{Paolo Politi}
\email{paolo.politi@isc.cnr.it}
\affiliation{Istituto dei Sistemi Complessi,
             Consiglio Nazionale delle Ricerche,
             via Madonna del Piano 10,
             I-50019 Sesto Fiorentino, Italy}
\affiliation{Istituto Nazionale di Fisica Nucleare, Sezione di Firenze,
             via G.~Sansone 1, I-50019 Sesto Fiorentino, Italy}
\author{Ruggero Vaia}
\affiliation{Istituto dei Sistemi Complessi,
             Consiglio Nazionale delle Ricerche,
             via Madonna del Piano 10,
             I-50019 Sesto Fiorentino, Italy}
\affiliation{Istituto Nazionale di Fisica Nucleare, Sezione di Firenze,
             via G.~Sansone 1, I-50019 Sesto Fiorentino, Italy}

\date{\today}

\begin{abstract}
Antiferromagnetic chains with an odd number of spins are known to
undergo a transition from an antiparallel to a spin-flop
configuration when subjected to an increasing magnetic field. We show
that in the presence of an anisotropy favoring alignment
perpendicular to the field, the spin-flop state appears for {\it
both} weak and strong field, the antiparallel state appearing for
intermediate fields. Both transitions are second order, the
configuration varying continuously with the field intensity. Such
re-entrant transition is robust with respect to quantum fluctuations
and it might be observed in different types of nanomagnets.
\end{abstract}



\maketitle

{\it Introduction ---}
 The ability to manipulate atoms adsorbed on a substrate,
\cite{HirjibehedinScience2006,LothScience2012,HolzbergerPRL2013}
the possibility to choose a suitable combination of substrate and adatoms,
and the recent capacity to tailor microscopic interactions
\cite{BezerraSR2013,BrovkoJPC2014}
permit to obtain nanosystems with specific magnetic properties.
Adatoms may interact either ferromagnetically or
antiferromagnetically; the coupling with a ferromagnetic substrate
may mimic an external field and it also induces spin anisotropies via
spin-orbit interaction. These couplings are the building blocks of a
variety of magnetic configurations, so it is not surprising that in
atomic chains one can recognize phenomena and transitions originally
discovered in bulk samples, then also studied in stratified systems.

The spin-flop transition is a well suited and important example of
magnetic phenomenon whose study has accompanied the race to
miniaturization. This transition is due to the competition between
antiferromagnetic coupling and magnetic field, and it connects an
antiferromagnetic configuration with spins aligned along the field to
a configuration where they are almost perpendicular to the field,
with a small tilting angle $\varepsilon$ producing a nonvanishing
magnetization. In the simplest case, an isotropic infinite system,
the critical field for such transition is zero, because the Zeeman
energy gain ($-H\varepsilon$) in tilting the angle dominates upon the
exchange energy loss ($J\varepsilon^2$) due to spin misalignment. On
the other hand, in the presence of a small easy-axis anisotropy
$\kappa<0$, a field applied along such axis must overcome a finite
critical threshold $H_c\approx{J}\sqrt{|\kappa|}$ in order to produce
the spin-flop reorientation. This transition is first order, the
system passing with discontinuity from the antiparallel (AP) to the
spin-flop (SF) phase. The metastability region has size $\delta H$,
with $\delta H/H_c\approx|\kappa|$.

When interfaces or finiteness are introduced, the magnetic phase
diagram becomes richer and new effects arise. An especially relevant
example is the parity
effect~\cite{WangPRL1994,TralloriIJMPB1996,LounisPRL2008,HolzbergerPRL2013}:
systems of different parity have different behaviors because an
antiferromagnetic (AFM) chain of $N$ spins has (odd $N$) or has not
(even $N$) a finite magnetization which couples to the external
field. In particular, for odd $N$ such residual magnetization may
stabilize the AP configuration with respect to the SF phase for
fields which can be much larger than
$H_c$~\cite{LounisPRL2008,PolitiPRB2009,BerzinPSS2010,Morosov2010,MattiaTesi2013}.
Here we show that the presence of an anisotropy favoring alignment
{\it perpendicular} to the field alters the above
scenario, determining two critical fields $H_\pm(N)$, with the AP
phase appearing for $H_-(N)<H<H_+(N)$ and the SF phase outside such
interval. For $N=N^*$, $H_-(N^*)=H_+(N^*)$ so that only the SF phase
appears for $N>N^*$. Furthermore, both transitions at $H_\pm(N)$ are
continuous.

\medskip

{\it Models and main results ---} We consider a chain with an {\em
odd} number $N\equiv2M{+}1$ of spins, described by the Heisenberg
Hamiltonian
\begin{equation}
 \hat{\cal H} = J \sum_{i=1}^{N-1}
 \hat{\mathbf{S}}_i{\cdot}\hat{\mathbf{S}}_{i+1}
  -H\sum_{i=1}^N \hat{S}^z_i + J\kappa\sum_{i=1}^N (\hat{S}^z_i)^2~.
\label{e.Hq}
\end{equation}
Special attention will be given to the easy-plane case ($\kappa>0$),
but we will also refer to $\kappa\le 0$.
Its classical counterpart (in units of $JS^2$) is given by
\begin{equation}
 E = \sum_{i=1}^{N-1} \cos(\theta_i{-}\theta_{i+1})
  -h\sum_{i=1}^N \cos\theta_i + \kappa \sum_{i=1}^N \cos^2\theta_i ~,
\label{e.Hc}
\end{equation}
where the azimuthal angles $\varphi_i$ have been taken equal, our
purpose being that of characterizing the minimum-energy state
(ground state); note that the physically relevant values of the
anisotropy are small, $|\kappa|\ll1$.

It is possible to qualitatively illustrate in simple terms the main
results that are more rigorously studied in the following Sections.
Assuming the SF configuration to be uniform, as in the infinite
system, $\theta_i=(-1)^i\theta$, from~\eqref{e.Hc} its classical
energy per site is about
$e\ss{SF}(\theta)\simeq\cos2\theta-h\cos\theta+\kappa\cos^2\theta$,
which is minimal for $2\cos\theta=h/(2{+}\kappa)$, giving
$e\ss{SF}\simeq-1{-}h^2/8$. As for the AP configuration,
$\theta_i=i\pi$, one easily finds $e\ss{AP}=-1{+}\kappa{-}h/N$, the
last term being due to the balance between $M{+}1$ up and $M$ down
spins, respectively. The condition $e\ss{AP}<e\ss{SF}$ reads
\begin{equation}
 \frac{h^2}8\,-\frac{h}N+\kappa<0 ~,
\label{e.eqh}
\end{equation}
implying that the AP phase is energetically favored for
\begin{equation}
 N <\, N_{\rm{c}}(h,\kappa)=\frac{h}{\kappa+h^2/8}~,
\label{e.Nc1}
\end{equation}
i.e., for $h$ between two `critical' field values
\begin{equation}
 h\in[h_-,h_+]~,~~~~
 h_{\pm}(N,\kappa)=\frac4N
  \bigg(1\pm\sqrt{1-\frac{\kappa\,N^2}2}~\bigg) ~.
\label{hpm}
\end{equation}
Hence, when rising the field from zero:
first, the system enters in the SF phase;
 at $h_-$ such phase is left and the AP one shows up;
finally, beyond $h_+$ the system re-enters the SF phase. Evidently,
this happens for small enough $N\lesssim\sqrt{2/\kappa}$, a value
beyond which the intermediate AP configuration disappears. It is
worth stressing that for a vanishing or easy-axis anisotropy
($\kappa\le0$), the lower `critical' field $h_-\le0$, meaning that at
small field the system is in the AP state for any $N$, and the only
transition at $h_+$ is left.

The above SF-AP-SF re-entrant transition should be accessible to
experiment (see the final Discussion). However, as real systems at a
few Kelvin are quantum mechanical, one has to account for the effect
of quantum fluctuations~(QFs). Any classically ordered state, such as
the AP one, is usually weakened by QFs. For instance, the quantum
three- and two-dimensional isotropic antiferromagnets are subject to
{\it spin reduction}~\cite{Kubo1952}, namely, QFs make the
ground-state sublattice magnetization smaller than $S$, while in the
one-dimensional case the ground state does not even show order, the
N\'eel AFM state being unstable under soft spin-wave excitations.
Therefore, it makes sense to ask whether QFs destroy the re-entrant
transition in the finite chain or not, a question that can be
answered by studying the stability of the AP state under QFs. A
na\"\i{ve} approach is to apply the renormalization
scheme~\cite{Koehler1966} known as self-consistent harmonic
approximation (SCHA) as done in
Refs.~\onlinecite{1992CTVVpra,1995CGTVVjpcm,1996CTVVprl,1997CTVVprb}
for a translation invariant two-dimensional AFM. In this approach,
zero-$T$ QFs can be substantially accounted for by the classical
system, but taking spins reduced by a factor $\alpha(S)=1{-}D$ (i.e.,
each quantum spin $\hat{\bm{S}}_i$ has a classical counterpart
$\alpha{\bm{S}}_i$). In the one-dimensional case
\begin{equation}
 D = \frac1{(2S{+}1)N}\sum_k\sqrt{1{-}\cos^2{k}}
   = \frac2{\pi(2S{+}1)} ~.
\end{equation}
It is then straightforward to see from Eq.~\eqref{e.Hq} that $J$
gains a factor $\alpha^2$, and $H$ a plain factor $\alpha$, so $h$ is
replaced by $h/\alpha$ in Eqs.~\eqref{e.Hc} and~\eqref{e.eqh}, and
eventually the quantum transition fields are expected at
$h_{\pm}^{\rm(q)}={\alpha}\,h_{\pm}(N,\kappa)$. According to this
argument the AP region should be maintained, but with smaller critical fields,
$h_{\pm}^{\rm(q)} < h_{\pm}$. However, as we will see in more detail later on,
this argument fails because it assumes homogeneous QFs. Instead, the
breaking of translation symmetry causes, in the vicinity of the chain
ends, smaller fluctuations of the odd spins (parallel to the field)
and larger ones for the even ones (antiparallel to the field). The
picture of `equal spin reduction' breaks down and the boundary spins,
which are less affected by QFs, remain more `resistant' towards the
incipient SF configuration. As a consequence, the stability of the AP
state is unexpectedly reinforced, and $h_+^{\rm(q)}$ becomes even
larger than $h_+(N,\kappa)$.

\medskip

{\it Classical model ---}
 Because of the broken translation invariance, minimum energy
configurations must be sought in the $N-$dimensional space
$\theta_1,\dots,\theta_N$, minimizing the energy \eqref{e.Hc}.
Defining $s_i=\sin(\theta_i{-}\theta_{i-1})$, the equations $\partial
E/\partial\theta_i=0$ can be cast in the form of a two-dimensional
mapping,\cite{TralloriIJMPB1996}
\begin{equation}
\left\{
\begin{split}
& s_{i+1} = s_i -h\,\sin\theta_i +\kappa\,\sin 2\theta_i \\
& \theta_{i+1} = \theta_i + \sin^{-1} s_{i+1} .
\end{split}
\right.
\label{e.map}
\end{equation}
The function $\sin^{-1}$ in \eqref{e.map} has two solutions; since
the AFM exchange coupling is strong, we must choose the solution such
that $(\theta_{i+1}-\theta_i) \in [\frac{\pi}{2},\frac{3\pi}{2}]$.
The absence of spins at $i=0$ and $i=N{+}1$ can be accounted by the
boundary conditions $s_1=s_{N+1}=0$. The angle $\theta_1$ is
determined imposing that after $N$ iterations we can satisfy the
condition $s_{N+1}=0$. It is worth remarking that the AP
configuration $\theta_i=i\pi$ is always a solution of
Eqs.~\eqref{e.map}. Its energy must be compared with possible
nonuniform solutions (SF) in order to identify the ground state.

\begin{figure}
\begin{center}
\includegraphics[height=80mm,angle=90]{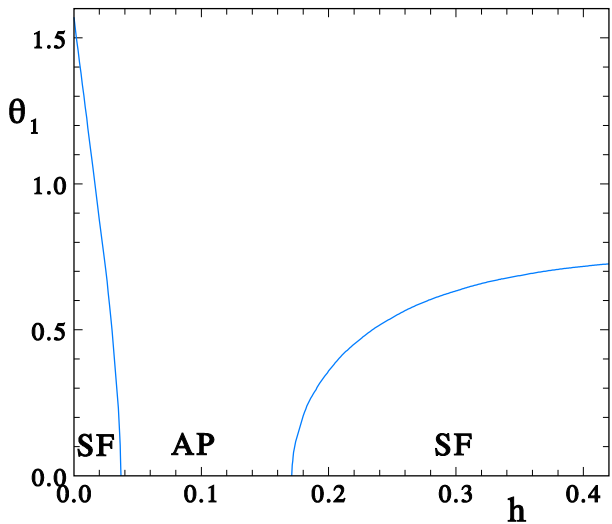}
\end{center}
\caption{
The angle $\theta_1$ corresponding to the classical ground-state
configuration as a function of $h$, for $N=15$ and $\kappa=0.001$. It
varies with continuity in the whole range of $h$. For large $h$ (not
shown) $\theta_1$ attains a maximum, then it decreases because the SF
phase gets ferromagnetic for $h\simeq 2(2+\kappa)$. }
\label{fig_theta1}
\end{figure}

\begin{figure}
\begin{center}
\includegraphics[height=80mm,angle=90]{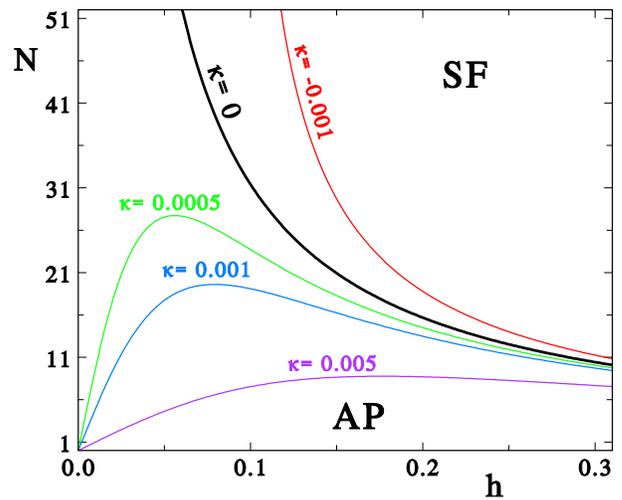}
\end{center}
\caption{
Phase diagram for a classical spin chain of length $N$ (odd), for easy-plane,
easy-axis, and no anisotropy.}
\label{f.classica}
\end{figure}

In Fig.~\ref{fig_theta1} we plot the angle between the first (or
last) spin and the field for $N<N^*$, so that two transitions are
crossed when increasing $h$. The continuous character of both
transitions will be discussed in the final Section. Generally
speaking, it is possible to numerically determine the phase diagram
in the $(h,N)$ plane for different values of $\kappa$ using the map
method, but this would be very lengthy. In
Ref.~\onlinecite{PolitiPRB2009} the limit of the AP phase has been
determined {\it analytically} when $\kappa=0$, the transition
corresponding to the vanishing of $s'_{N+1}(\theta_1)$ in
$\theta_1=0$. In fact, it is possible to extend the same
procedure~\cite{unpublished} to the anisotropic case, obtaining the
curves shown in Fig.~\ref{f.classica}. We have also derived the
analytical phase boundaries by studying the stability of the AP
phase, which is also needed for the quantum treatment explained
below. In the classical limit this approach yields the very same
results obtained by the map
analysis~\cite{PolitiPRB2009,unpublished}. In view of the quantum
treatment it is useful to extend the model to include an exchange
anisotropy, described by adding to Eq.~\eqref{e.Hq} the term
\begin{equation}
  -J\lambda \sum_{i=1}^{N-1} \hat S^z_i \hat S^z_{i+1}~.
\label{e.Hqq}
\end{equation}
Such a different anisotropy makes our results more general and allows
us to show that they do not depend on the details of the anisotropy,
but only on its sign. The exact result for the phase-boundary is
\begin{equation}
  N_{\rm{c}}(h,\kappa,\lambda)
  =\frac {\displaystyle \tan^{-1}\bigg[
  \frac12\,\frac{h{+}2\lambda\mu}
  {\kappa{+}\lambda\mu}
  ~a(h,\kappa{+}\lambda)\bigg]}
  {\displaystyle \tan^{-1}\!~a(h,\kappa{+}\lambda)} ~,
\label{e.Nc}
\end{equation}
where $\tan^{-1}$ is defined with the codomain $[0,\pi]$,
$\mu=(1{-}\lambda/2)(1{-}\kappa{-}\lambda{-}h/2)$, and
\begin{equation}
 a(h,x)=\sqrt{\big[(1{-}x)^2{-}\,h^2/4\big]^{-1}-1}~.
\end{equation}
When $\kappa$ and $\lambda$ are small, $\mu\simeq1$, so that
$N_{\rm{c}}$ only depends on the sum $\kappa{+}\lambda$, i.e.,
exchange- and single-site anisotropies are almost equivalent in the
classical system.

In Fig.~\ref{f.classica} we report the results of the above analysis
for three cases: easy-axis anisotropy ($\kappa<0$), no anisotropy
($\kappa=0$), and easy-plane anisotropy ($\kappa>0$). All the curves
are qualitatively reproduced by Eq.~\eqref{hpm}: for $\kappa>0$ the
AP state exists for $h_-<h<h_+$ up to a value $N^*$ of the chain
length; for $\kappa\le 0$ such state exists for $0<h<h_+$, with
$h_+(N)$ vanishing at large $N$ for $\kappa=0$ and going to the limit
$h_+(\infty)\approx\sqrt{|\kappa|}$ for $\kappa<0$.

\begin{figure}
\begin{center}
\includegraphics[height=80mm,angle=90]{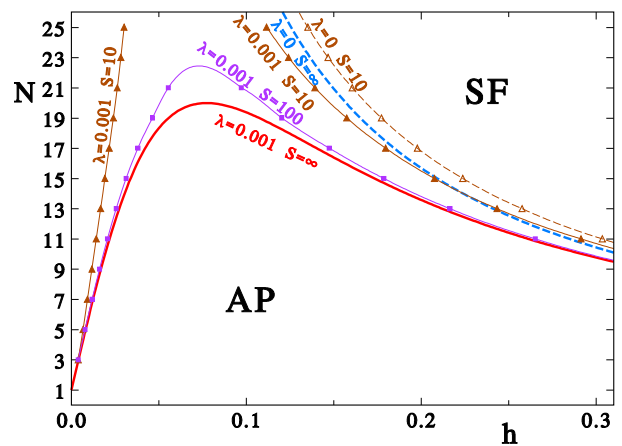}
\end{center}
\caption{
Phase diagram for a quantum spin chain of length $N$ (odd), for
easy-plane exchange anisotropy $\lambda=0.001$ and no anisotropy
($\lambda=0$) and for selected spin values.}
\label{f.quantum}
\end{figure}

\medskip

{\it Quantum model ---}
 In order to estimate the effects on the finite quantum chain
one has to account for the lack of translation symmetry and for the
different behavior of the two sublattices by means of a more accurate
quantum SCHA. The strategy is to study the linear excitations in the
AP ordered phase, which is assumed to be stable if the corresponding
frequencies are positive: at the phase boundary at least one
frequency vanishes, so that calculating the quantum-renormalized mode
frequencies allows us to draw conclusions for the quantum phase
diagram.

We restrict the quantum approach to the case of exchange anisotropy,
which can be treated unambiguously with respect to single-site
anisotropy: for instance, the latter is completely ineffective for
$S=1/2$, as $(\hat{S}^z)^2=1/4$. Therefore, we consider the
Hamiltonian~\eqref{e.Hq} with $\kappa=0$ and with the additional
term~\eqref{e.Hqq}. Performing the canonical transformation
$(\hat{S}^x,\hat{S}^y,\hat{S}^z)\longrightarrow
(-\hat{S}^x,\hat{S}^y,-\hat{S}^z)$ on the even indexed sites we get
the following Hamiltonian
\[
 \hat{\cal H}
 = -J\!\sum_{i=1}^{N-1}\big(
     \hat S_i^x\hat S_{i+1}^x{-}\hat S_i^y\hat S_{i+1}^y
      {+}\lambda\hat S_i^z\hat S_{i+1}^z\big)
         + H \sum_{i=1}^N (-)^i\hat S_i^z.
\]
The AP state corresponds to the fully aligned (N\'eel) state:
evidently, it is not an eigenstate of $\hat{\cal{H}}$. However, the
exact ground state is expected to be `close' to it, as it happens
with the ground state of bulk antiferromagnets; in the latter case,
the ground state and its linear excitations are found by means of a
Bogoliubov transformation for the Fourier-transformed operators. But
in our problem there is no translation symmetry, so the procedure is
necessarily more involute.

To proceed, we use the magnon creation and annihilation operators
defined by the Holstein-Primakoff~\cite{HolsteinP1940}
transformation:
$\hat{S}^x{+}i\hat{S}^y=\sqrt{2S{-}\hat{a}^\dagger\hat{a}}~\hat{a}$,
$\hat{S}^z=S{-}\hat{a}^\dagger\hat{a}$. Then, we introduce
coordinates and momenta defined by
$\hat{q}_i=(\hat{a}_i^\dagger{+}\hat{a}_i)/\sqrt{2S{+}1}$ and
$\hat{p}_i=i(\hat{a}_i^\dagger{-}\hat{a}_i)/\sqrt{2S{+}1}$, in such a
way that $[\hat{q}_i,\hat{p}_j]=i\delta_{ij}(S{+}\frac12)^{-1}$ (the
classical limit occurring for $S\,{\to}\,\infty$). It is easier to
work with their Weyl symbols~\cite{Berezin1980} $q_i$ and $p_i$, so
complex quantities do not appear, and expanding up to quartic
terms~\cite{1997CTVVprb} one is left with a quadratic Hamiltonian and
a quartic interaction,
\begin{eqnarray*}
 {\cal H}_2 &=& \sum_{i=1}^{N-1}\big[
      p_ip_{i+1}{-}q_iq_{i+1}+\lambda(z_i{+}z_{i+1}) \big]
      - h \sum_{i=1}^N (-)^i z_i ~,
\\
 {\cal H}_4 &=& \sum_{i=1}^{N-1}\big[
      {\textstyle\frac14}(z_i{+}z_{i+1}) (q_iq_{i+1}{-}p_ip_{i+1})
      -\lambda z_iz_{i+1} \big] ~,
\end{eqnarray*}
where $z_i\equiv({p_i^2}{+}q_i^2)/2$. Note that
${\cal{H}}_2=\frac12~\big(\bm{p}^t\bm{A}^2\bm{p}
+\bm{q}^t\bm{B}^2\bm{q}\big)$ is a quadratic form with
$[\bm{A}^2,\bm{B}^2]\ne{0}$, so its reduction to independent normal
modes~\cite{Colpa1978,Tsallis1978,Arnold1978} is nontrivial. First,
one has to assume the $N{\times}N$ matrices $\bm{A}^2$ and $\bm{B}^2$
to be positive definite: thanks to their structure,
this can be assessed analytically, and gives the classical AP phase
boundary mentioned above. Then, one performs the canonical
transformation
$(\bm{q},\bm{p})\longrightarrow(\bm{A}\bm{q},\bm{A}^{-1}\bm{p})$, so
that ${\cal{H}}_2\longrightarrow\frac12~\big(\bm{p}^t\bm{p}
+\bm{q}^t\bm{A}\bm{B}^2\bm{A}\bm{q}\big)$ and eventually the
eigenfrequencies $\omega_k^2$ arise as eigenvalues of
$\bm{A}\bm{B}^2\bm{A}$, while the decoupled modes have the
ground-state QFs $\langle{p}^2_k\rangle=\omega_k/(2S{+}1)$ and
$\langle{q}^2_k\rangle=1/[\omega_k(2S{+}1)]$. Renormalizations come
into play through ${\cal{H}}_4$, which is treated within the
SCHA~\cite{Koehler1966}, yielding for $\bm{A}^2$ and $\bm{B}^2$
corrections which self-consistently depend on the correlators
generated by ${\cal{H}}_2$. Such a perturbative approximation is the
more reliable the smaller the corrections it gives. For systems in
three and two dimensions the high coordination degree yields small
fluctuations, which only slightly modify the ground state and the low
temperature phase, while in the present quasi-one-dimensional system
the matrix elements of $\bm{A}^2$ and $\bm{B}^2$ are corrected by
terms of order $1/S$ which become rapidly large when considering
small spin values; however, if one `switches on quanticity' starting
from large spin values, the trend towards an extension of the AP
region is evident. The above procedure can be easily implemented
numerically and the final quantum results are reported in
Fig.~\ref{f.quantum}, which confirms the overall classical scenario
of Fig.~\ref{f.classica}.

\medskip

{\it Discussion ---} We have proven that an AFM chain composed by an
odd number of spins undergoes an unusual re-entrant spin-flop
transition, if the magnetic field is applied along a hard axis. This
result appears to be rather robust: the details of the anisotropy
and the classical or quantum character of the model are irrelevant.
In fact, the qualitative explanation of the double transition we have
given in Eqs.~\eqref{e.eqh}-\eqref{hpm} is fairly simple and the
subsequent Sections confirm such a result. It is worth stressing
that, at least classically, the ground state of the chain in the
presence of an easy-plane anisotropy favoring the $\hat{xy}$ plane
($\kappa>0$) does not change if the anisotropy is easy-axis and
favors any direction in such plane.~\footnote{Spin-wave excitations
and zero-point fluctuations make the two models different, because
their symmetry properties are different, but their classical ground
states are the same.}

Figure~\ref{fig_theta1} shows that $\theta_1$ varies with continuity
with $h$, making second order the SF-AP (at $h=h_-$) and the AP-SF
(at $h=h_+$) transitions. Such character is in contrast with the
standard spin-flop transition appearing when the field is applied
along an easy axis ($\kappa < 0$). In this case, the infinite system
undergoes a first order transition at
$h_c=2\sqrt{|\kappa|(2-|\kappa|)}$, with a metastability region of
size $\delta h\approx |\kappa|^{3/2}$. This behavior is maintained
for finite systems if (odd) $N$ is large enough. When $\kappa=0$,
$h_c=0$ and the transition gets continuous, which is not surprising,
since the metastability region is negligibly small with respect to
$h_c$, when $\kappa\to 0^-$. The continuous character at $\kappa=0$
is preserved when $\kappa>0$, as we have verified for $N=15$
(Fig.~\ref{fig_theta1}) and for large $N$ (not shown).

Our models, either classical, Eq.~\eqref{e.Hc}, or quantum,
Eq.~\eqref{e.Hq}, have several possible experimental counterparts. We
cite here the most relevant two, namely layered systems and chains of
magnetic adatoms on a magnetic substrate. A superlattice A/B, if one
of the two materials (A) is ferromagnetic and the indirect coupling
between A layers, mediated by the spacer B is antiferromagnetic, can
be studied by our model Eq.~\eqref{e.Hc},
where each spin represents the effective magnetization of a single A
layer (quantum effects are therefore negligible, since $S$ is macroscopic and
the temperature is high). In fact, Fe/Cr(211) superlattices have been
used~\cite{WangPRL1994} to study surface effects on the spin-flop
transition, but the field was applied in an easy direction
($\kappa<0$).
So, parity effects were visible, but the distinctive spin-flop
transitions appearing for odd $N$ when the field is along an hard
axis were overlooked.

In the last few years, much experimental effort has been devoted to
completely different magnetic systems undergoing a spin-flop
transition: magnetic atoms deposited on a substrate and forming a
linear chain. An existing experimental set-up which is appropriate
for our study is a recent one~\cite{HolzbergerPRL2013}, where Mn
adatoms were deposited and manipulated on top of a Ni(110) surface.
In that paper authors compare successfully experiment and theory for
the ground states of a dimer ($N=2$) and a trimer ($N=3$). For $N=3$
they actually find an AP state, which is in agreement with our phase
diagram if we use their experimental values $h\simeq 0.5$ and
$\kappa\approx 0.001$. The limit of using adatom chains for studying
SF transitions is that the field is fixed, because it is not a true
magnetic, external field. On the contrary, changing $N$ is much
easier than for layered systems. However, recent experimental results
suggest that tuning the parameters of Eqs.~\eqref{e.Hq}-\eqref{e.Hc}
is indeed possible~\cite{BezerraSR2013,BrovkoJPC2014}.

Our results might also be relevant for a third class of nanosystems,
namely molecular
nanomagnets~\cite{Gatteschi2006,Micotti2006,Ochsenbein2008}. In this
case, the coupling between spins is weaker and an external magnetic
field can be applied to tune the configuration. In
conclusion,  the SF-AP-SF re-entrant transition we have discussed in
this paper might be really accessible to experiments.

\medskip

We thank Gloria Pini for useful discussions and Wulf Wulfhekel for
pointing out some relevant references.

\end{document}